\newcommand{\rem}[1]{}
\documentclass[11pt]{amsart}
\usepackage{url}
\usepackage{epstopdf}
\newtheorem*{theorem*}{Theorem}
\newtheorem*{remark*}{Remark}
\newtheorem{thrm}{Theorem}[section]
\newtheorem{lem}[thrm]{Lemma}
\newtheorem{prop}[thrm]{Proposition}

\theoremstyle{definition}

\begin{document}
\author[C.~A.~Mantica, A.~Tacchini, L.~G.~Molinari,]
{Luca~Guido~Molinari, Adriano Tacchini \\and Carlo~Alberto~Mantica}
\address{L.~G.~Molinari and A.~Tacchini: Physics Department,
Universit\`a degli Studi di Milano and I.N.F.N. sez. di Milano,
Via Celoria 16, 20133 Milano 
-- C.~A.~Mantica: I.I.S. Lagrange, Via L. Modignani 65, 
20161, Milano, Italy and I.N.F.N. sezione di Milano}
\email{luca.molinari@unimi.it, adriano.tacchini@studenti.unimi.it} 
\email{carlo.mantica@mi.infn.it}
\date{04 august 2019}
\begin{abstract} 
We prove that in space-times a velocity field that is shear, vorticity and accelera\-tion-free, if any,  is unique 
up to reflection, with these exceptions: generalized Robertson-Walker space-times
whose space sub-manifold is warped, and
 twisted space-times (the scale function is space-time dependent)
whose space sub-manifold is doubly twisted. In space-time dimension $n=4$,
the Ricci and the Weyl tensors are specified, and the Einstein equations yield
a mixture of two perfect fluids. 
\end{abstract}
\title[On uniqueness of torse-forming velocity]{On the uniqueness of a\\
shear-vorticity-acceleration-free \\
velocity field in  space-times}

\maketitle

\section{Introduction and statement of results}
In a space-time of dimension $n>3$, let $u_k$ be a smooth velocity field that is shear-free, vorticity-free
and acceleration-free:
\begin{align}
u^ku_k =-1, \qquad \nabla_i u_j = \varphi (u_iu_j + g_{ij}), \label{torse}
\end{align}
in other words, $u_k$ is a time-like unit torse-forming vector field, with scalar field $\varphi $. 
We enquire whether the space-time may admit other velocity fields that are  
time-like unit and torse-forming, 
\begin{align}
w^kw_k=-1,\qquad \nabla_iw_j = \lambda (w_iw_j + g_{ij}) \label{omega}
\end{align}
besides the trivial twin vector $-u_k$ with scalar field $-\varphi $.

The existence of the vector field $u_k$ ensures that the space-time is Twisted \cite{ManMol2017}, 
i.e. there is a reference frame where the metric has the form:
\begin{align}
 ds^2 =- dt^2 + a(t,x)^2 g^*_{\mu\nu} (x) dx^\mu dx^\nu  \label{TW}
 \end{align}
where $t$ is the time, the scale function $a(t,x)>0$ depends on time and space coordinates, and $g^*_{\mu\nu}(x)$ is the metric tensor of a Riemannian 
sub-manifold $M^*$ with space coordinates $x^\mu$. Twisted space-times were introduced by
B.-Y. Chen, to generalise the notion of warped manifolds \cite{Chen1979}\cite{Chen2017}.\\
In the locally ``comoving'' frame \eqref{TW} it is $u^0=1,\, u^\mu=0$ and $\varphi =\dot a/a$.
With the Christoffel symbols listed in the Appendix, the normalisation and torse-forming conditions for another
vector $w_i$ are the equations:
\begin{align}
& -w_0^2 +a^{-2}g^{*\mu\nu} {w_\mu}{w_\nu} = -1, \quad\quad \partial_t w_0 = \lambda (w^2_0 -1),\nonumber \\
&\partial_\mu w_0 - (\dot a/a){w_\mu} = \lambda w_0{w_\mu} , \quad 
\partial_t{w_\mu} -(\dot a/a) {w_\mu}  =\lambda w_0{w_\mu}  \label{eqT}\\
&\nabla^*_\mu {w_\nu} -\frac{a_\mu}{a}{w_\nu} -
\frac{a_\nu}{a}{w_\mu} + g^*_{\mu\nu}(\frac{a_\rho}{a}w_\sigma g^{*\rho\sigma}-
a\dot a w_0)= \lambda ({w_\mu}{w_\nu} + a^2 g^*_{\mu\nu})  \nonumber
\end{align}
where $\dot a=\partial_t a$, $a_\mu =\partial_\mu a$, and $\nabla^*$ is the connection of the sub-space ($M^*,g^*$). 
To investigate the existence of a non-trivial solution (besides $u_0=-1$, $u_\mu=0$, $\lambda =\varphi $) we distinguish three cases:
\begin{align*}
\text{Case A)}\quad & \nabla_k\varphi =0 \;  \text{and $\varphi \neq 0$},\;\text{($\varphi =0$ factors space and time)};  \\
\text{Case B)}\quad &  \nabla_k\varphi =-u_k u^m\nabla_m \varphi \,;\\
\text{Case C)}\quad & \nabla_k\varphi =-u_k u^m\nabla_m \varphi + v b_k, \;\; \text{with $v\neq 0,\; b_kb^k=1,\;
u^kb_k=0$}.
\end{align*}
\noindent
In Cases A and B the existence of $u_k$ implies that the space-time is warped, i.e. there is a reference frame
where the metric \eqref{TW} has the simpler form 
\begin{align}
ds^2 =- dt^2 + a(t)^2 g^*_{\mu\nu} (x) dx^\mu dx^\nu \label{warped}
 \end{align} 
where the scale function only depends on the time. Such space-times are also named 
generalised Robertson-Walker (GRW) (see \cite{ManMolGRW} for a review,
and \cite{Chen2014} for another covariant characterisation).
They are Robertson-Walker (RW) if the Weyl tensor is zero. 

For each case we prove:\\

{\bf Theorem A.} {\em In a warped space-time \eqref{warped} with constant non-zero $\varphi $, the torse-forming velocity is unique unless  
$(M^*,g^*)$ is a warped sub-manilfold, i.e. ($M^*,g^*$) admits a unit vector field $n_\mu^*(x)$ such that 
$\nabla^*_\mu n^*_\nu = \theta (x) (n^*_\mu n^*_\nu - g^*_{\mu\nu} ) $ with $\nabla_\mu^* \theta $  proportional
to $n^*_\mu$.}\\

{\bf Theorem B.} {\em In a warped space-time \eqref{warped} with non-constant $\varphi $, the torse-forming velocity is unique.}\\

{\bf  Theorem C.} {\em
In a twisted space-time \eqref{TW} 
the torse-forming
velocity is unique unless  $(M^*,g^*)$ is a doubly-twisted sub-manifold. This is equivalent to the requirement
\begin{align}
\nabla_ib_j =\varphi b_i u_j +(g_{ij}+u_iu_j -b_ib_j)\frac{\nabla_kb^k}{n-2} 
\label{condition}
\end{align}
The second torse-forming velocity is  $w_i=u_i\cosh \alpha + b_i \sinh\alpha $ with:}
\begin{align}
& \tanh\alpha = - \frac{2R_{ij}u^ib^j}{R_{ij}(u^iu^j+b^ib^j)}\\ 
& \lambda = \varphi \cosh\alpha + \frac{\nabla_kb^k}{n-2} \label{lambda}
 \end{align}
The property $|\tanh\alpha |\le 1$ poses a restriction.

Theorems A, B and C correct Prop.2.3 in our paper \cite{ManMol2017} that claimed uniqueness in all cases. In its short proof, the sign in front of $(\dot f/f)\sinh a V_\mu $ in the centred equation is wrong. With the correct (minus) sign, the proof
by reductio ad absurdum does not work. The error does not affect the rest of the paper.

Cases A, B and C will be discussed separately, case B being simpler and preparatory for case A. For all cases,
some preliminary identities that simplify the discussion are first obtained. 

Lovelock's identity in $n=4$ and the existence of a second torse-forming vector, determine the electric component of the Weyl tensor and thus the Ricci and Weyl tensors. By the Einstein equations, an energy-momentum
tensor $T_{ij}$ is obtained, that describes a mixture of two perfect fluids with non collinear velocities, studied by Coley and McManus \cite{ColeyMcM}\cite{ColeyMcM96}. In their work, the request that one velocity is torse-forming (no restriction 
for the other velocity),  implies that the subspace $M^*$ admits an umbilical foliation i.e. it is doubly twisted \cite{Ponge}\cite{Chen2017}. In the present study, the mixture of perfect fluids arises in $n=4$ as a consequence of having two torse-forming vectors and only one of them, in general, is the velocity of one of the fluids.

\section{Preliminary results}
Suppose that, besides $u_k$ with scalar field $\varphi $, there exists another time-like unit torse-forming vector field $w_k$ with scalar field $\lambda $, eq.\eqref{omega}, not collinear with $u_k$.
\begin{remark*}
We are assuming $(u^kw_k)^2\neq 1$, otherwise $w_k$ would be space-like.
\end{remark*}
\noindent
The following identities with the Riemann tensor, $[\nabla_i,\nabla_j]u_k =R_{ijkm}u^m$ and $[\nabla_i,\nabla_j]w_k =R_{ijkm}w^m$, are
evaluated with \eqref{torse} and \eqref{omega}:
\begin{align}
&R_{ijkm} u^m = (u_ju_k +g_{jk})\nabla_i\varphi -(u_iu_k + g_{ik} ) \nabla_j\varphi +\varphi^2 (u_j g_{ik}-u_i g_{jk}) \label{nabnab} \\
&R_{ijkm} w^m = (w_jw_k +g_{jk})\nabla_i\lambda -(w_iw_k + g_{ik} ) \nabla_j\lambda +\lambda^2 (w_j g_{ik}-w_i g_{jk}) \label{nabnab2}
\end{align}
Contraction of both equations with $g^{ik}$ gives identities with the Ricci tensor:
\begin{align}
&R_{jm} u^m = (u_ju_k +g_{jk})\nabla^k\varphi +(n-1)(\varphi^2 u_j -\nabla_j\varphi ) \label{ricci1} \\
&R_{jm} w^m = (w_jw_k +g_{jk})\nabla^k\lambda +(n-1)(\lambda^2 w_j- \nabla_j\lambda) \label{ricci2}
\end{align}
Transvect \eqref{nabnab} with $w^i$ and \eqref{nabnab2} with $u^i$. In the second one use the symmetries of the Riemann tensor:
$R_{ijkm}u^i w^m=
R_{mkji} u^iw^m=R_{ikjm}w^iu^m$, then exchange $k$ with $j$:
\begin{align*}
&R_{ijkm}w^i u^m = (u_ju_k +g_{jk})w^i\nabla_i\varphi -(w^i u_iu_k + w_k ) \nabla_j\varphi +\varphi^2 (u_j w_k -w^i u_i g_{jk})\\ 
&R_{ijkm} u^mw^i = (w_jw_k +g_{jk})u^i\nabla_i\lambda -(u^iw_iw_j + u_j) \nabla_k\lambda +\lambda^2 (w_k u_j-u^iw_i g_{jk}). 
\end{align*}
Subtract one equation from the other:
\begin{align*}
g_{jk}[w^i\nabla_i\varphi -(\varphi^2-\lambda^2) w^iu_i -u^i\nabla_i \lambda]  +u_ju_k (w^i\nabla_i\varphi ) 
-(w^i u_iu_k + w_k ) \nabla_j\varphi \\
+(\varphi^2-\lambda^2) u_j w_k  
- w_jw_k (u^i\nabla_i\lambda )+(u^iw_iw_j + u_j) \nabla_k\lambda =0. \nonumber
\end{align*}
Contraction with a non-zero vector orthogonal to $u_j$, $w_j $ and $\nabla_j\varphi $ gives the equation
\begin{align}
w^i\nabla_i\varphi  - u^i\nabla_i \lambda =  (u^iw_i )(\varphi^2 - \lambda^2)  \label{eq_12a}
\end{align}
and, after simplification, the following one:
\begin{align}
u_ju_k (w^i\nabla_i\varphi )
-(w^i u_iu_k + w_k ) \nabla_j\varphi +(\varphi^2-\lambda^2) u_j w_k \label{eq_12b}\\
 - w_jw_k (u^i\nabla_i\lambda  ) 
+(u^iw_iw_j + u_j) \nabla_k\lambda =0.  \nonumber 
\end{align}
The trace of the latter is:
$(w^i u_i) (\varphi^2-\lambda^2 +w^j\nabla_j\lambda -u^j \nabla_j\varphi ) 
+ 2(u^i\nabla_i\lambda  -w^j\nabla_j \varphi )=0 $, with the aid of eq. \eqref{eq_12a} we obtain, after cancellation of $u^jw_j\neq 0$:
\begin{align}
 u^j \nabla_j\varphi +\varphi^2 = w^j\nabla_j\lambda + \lambda^2   \label{eq_13}
\end{align}
Hereafter we denote 
\begin{align}
\xi = (n-1)(u^j \nabla_j\varphi +\varphi^2)
\end{align}
Contraction of \eqref{eq_12b} with $u^j$ or with $w^k$ and use of \eqref{eq_13}, give:
\begin{align}
u_k [w^i\nabla_i\varphi +
w^i u_i (u^j \nabla_j\varphi) ] +w_k [w^j\nabla_j\lambda + u^iw_i (u^j\nabla_j\lambda  )]
=[(u^iw_i)^2 -1] \nabla_k\lambda   \label{13a} \\
u_j [u^i\nabla_i\varphi +
w^k u_k (w^i \nabla_i\varphi) ] +w_j [u^i\nabla_i\lambda + u^iw_i (w^k\nabla_k\lambda  )]
=[(u^iw_i)^2 -1] \nabla_j\varphi   \label{13b} 
\end{align}

\section{Proof of Theorem B}\label{sec_B}
\begin{lem}
If $\nabla_i\varphi = -u_i u^k\nabla_k\varphi $,  then $\nabla_i\lambda = -w_iw^k\nabla_k\lambda $.
\begin{proof}
Eq.\eqref{13a} simplifies to
$ w_k [w^j\nabla_j\lambda + u^iw_i (u^j\nabla_j\lambda  )]
=[(u^iw_i)^2 -1] \nabla_k\lambda $, showing that $\nabla_k\lambda $ is collinear to
$w_k$. It follows that $\nabla_k \lambda = - w_k (w^j \nabla_j \lambda )$. 
\end{proof}
\end{lem}
This result simplifies eqs.\eqref{ricci1} and \eqref{ricci2}, showing that $u_k$ and $w_k$ are both eigenvectors of the Ricci tensor with eigenvalue $\xi $:
$$R_{ij}u^j =\xi u_i, \qquad R_{ij}w^j =\xi w_i$$ 
Now, the problem is about degeneracy of the eigenvalue of the Ricci tensor:

\begin{prop}\label{degen} In a warped space-time, for the eigenvalue $\xi $ of the Ricci tensor to be degenerate, it is necessary that $\dot a(t)/a(t) = A t+B $, 
with constants $A $ and $B $.
\begin{proof}
Let us consider the eigenvalue equation $R_{ij}w^j = \xi w_i $ in the warped frame \eqref{warped}. The components of the Ricci tensor can be read in \cite{ManMol2017}.\\
The equation $ R_{00}w^0 = \xi w_0 $ is $-(n-1)(\ddot a/a)w^0 =\xi w_0$, then $\xi = (n-1)(\ddot a/a)$ as $w_0\neq 0$. 
In the equation $R_{\mu 0}w^0 + R_{\mu \nu}w^\nu = \xi{w_\mu} $
 one has $R_{\mu 0}=0$ and $R_{\mu \nu}= R^*_{\mu\nu}+g^*_{\mu\nu}[(n-2)\dot a^2 + a\ddot a] $. A solution is always $u^0=1$, $u^\mu =0$ (the given vector).
Other solutions have non-zero space components $w^\mu $ solving the eigenvalue equation:
\begin{align*}
R^*_{\mu\nu} w^\nu = &\xi {w_\mu} - [(n-2)\dot a^2 + a\ddot a] g^*_{\mu\nu}w^\nu \\
=& (n-2) \left(\frac{d}{dt}\frac{\dot a}{a}\right ) {w_\mu}
\end{align*}
where we lowered an index: $a^2 g^*_{\mu\nu}w^\nu = {w_\mu} $. In the warped frame the Ricci tensor $R^*_{\mu\nu}$ of $M^*$ does
not depend on time, and so must the eigenvalue. Then $(\dot a/a)=A t+B $ where $A$ and $B$ do not depend on space
coordinates, as the warping function does not.
\end{proof}
\end{prop}

\begin{lem}
If $\nabla_i\varphi = -u_i u^k\nabla_k\varphi $  then: $\nabla_k \xi = -u_k (u^j\nabla_j \xi )$.
\begin{proof} Evaluate: $\nabla_k \varphi^2 = 2\varphi \nabla_k\varphi = 2\varphi (-u_k u^j\nabla_j \varphi ) 
= -u_k u^j\nabla_j\varphi^2 $. Next:\\
$\nabla_k (u^j\nabla_j \varphi ) = \varphi (u_ku^j+\delta_k{}^j)\nabla_j\varphi + u^j\nabla_j\nabla_k\varphi $;
the first term is zero, the second one is: $u^j\nabla_j (-u_k u^\ell\nabla_\ell \varphi ) = 
-u_k  [ u^j\nabla_j (u^\ell\nabla_\ell \varphi )]$.  Add results and multiply by $(n-1)$.
\end{proof}
\end{lem}
\noindent
The same assertion holds for the torse-forming velocity $w_k$: $\nabla_k \xi = - w_k w^j \nabla_j \xi $. Comparison of assertions gives: if $\xi $ is not a constant scalar, then the torse-forming velocities 
$u_j$ and $w_j$ are collinear i.e. $u_j$ is unique. 

What remains to discuss is the case that $\xi $ is a constant scalar and is a degenerate eigenvalue. Proposition \ref{degen} imposes 
$\varphi =A t + B $ i.e. $\xi = (n-1)[(A t+B)^2+A]$. Then $\xi $ is constant if $A =0$ i.e. $\varphi  $ is constant, which is case A. \\
This proves Theorem B.

\section{Proof of theorem A}\label{sec_A}
\begin{lem} If $\varphi $ is a non-zero constant, and if $w_k$ exists not parallel to $u_k$, then $\lambda =\varphi$.
\begin{proof}
If $\lambda $ is constant, eq.\eqref{eq_13} implies $\lambda =\varphi $.\\
Now suppose that $\lambda $ is not a constant. Eq.\eqref{13a} with constant $\varphi $ gives: $w_k [w^j\nabla_j\lambda + u^iw_i (u^j\nabla_j\lambda  )]=[(u^iw_i)^2 -1] \nabla_k\lambda $. Then 
 $\nabla_k\lambda $ is proportional to $w_k$ i.e. $\nabla_k\lambda = -w_kw^j\nabla_j \lambda $.\\
Then both $u_j$ and $w_j$ are eigenvectors of the Ricci tensor with the same eigenvalue $\xi =(n-1)\varphi^2$. 
Given $w_j$ not collinear with $u_j$, there is a warped frame where $w^0=1$, 
$w^\mu =0$, and scale factor $\tilde a(t)$ such that $\dot {\tilde a}/\tilde a=\lambda $. As in Proposition \ref{degen} the condition that $\xi $ is degenerate and constant puts $\dot {\tilde a}/{\tilde a}=\lambda $ constant in space-time, and this
is against the hypothesis.
\end{proof}
\end{lem}

Being $\dot a/a=\varphi $ a non-zero constant, we can set $a(t) = \exp (\varphi t)/\varphi $.
The torse-forming conditions \eqref{eqT} simplify 
\begin{align*}
&\partial_t w_0 = \varphi (w_0^2-1)\quad
\partial_\mu w_0 = \varphi {w_\mu}(w_0+1)\quad
\partial_0{w_\mu}  = \varphi {w_\mu}(w_0+1)\\
&\nabla^*_\mu {w_\nu}   = \varphi [{w_\mu}{w_\nu} + a(t)^2 (w_0+1) g^*_{\mu\nu} ]
\end{align*}
The first three equations are solved by
$$ w_0(x,t) = \frac{1+C^2(x)\exp(2\varphi t)}{1-C^2(x)\exp (2\varphi t)}, \qquad 
{w_\mu} (x,t) = \frac{1}{\varphi }\frac{(\partial_\mu C^2)
\exp(2\varphi t)}{1-C^2(x)\exp (2\varphi t)} $$
where the function $C^2(x)$ is determined by the last differential equation:
$$ \partial_\mu\partial_\nu C^2(x) -\Gamma_{\mu\nu}^{*\rho} \partial_\rho C^2(x) =2 g^*_{\mu\nu} (x) \quad i.e.\quad
\nabla^*_\mu \nabla^*_\nu C^2 = 2 g^*_{\mu\nu} $$
The normalization condition $-1=-w_0^2+\varphi^2 e^{-2\varphi t} g^{*\mu\nu} {w_\mu}{w_\nu} $ gives:
$$4C^2 = g^{*\mu\nu}(\partial_\mu C^2)( \partial_\nu C^2 )  $$
If we put $\partial_\mu C=n^*_\mu $, then:
$g^{*\mu\nu}n^*_\mu n^*_\nu =1$ and $\nabla^*_\mu n^*_\nu =  \frac{1}{C} (g^*_{\mu\nu}-n^*_\mu n^*_\nu) $
i.e. $n^*_\mu $ is unit and torse-forming in $(M^*,g^*)$. Since also
$\partial_\mu (1/C) = -n^*_\mu/C^2$ the Riemannian
subspace ($M^*,g^*$) is warped, i.e. there is a choice of
space coordinates such that 
$ g^*_{\mu\nu} dx^\mu dx^\nu  =  (dx^1)^2 + f^2(x^1)ds^2$, where $ds^2$ involves the coordinates  $x^2,...,x^{n-1}$.


\section{Proof of theorem C}\label{sec_C}
If $\nabla_k \varphi $ is not collinear with $u_k$, the coefficient of $w_k $ in eq.\eqref{13b} 
cannot be zero, and the same equation shows that $\nabla_k \varphi$
is a linear combination of $u_k$ and $w_k $.
Eq.\eqref{13a} shows that also $\nabla_k\lambda $ 
is a linear combination of $u_k$ and $w_k$.\\
If $\nabla_k\varphi +u_k (u^i\nabla_i\varphi ) = v b_k $, where $v\neq 0$, $b^jb_j=1$, $b_j u^j=0$, then $w_k$ is spanned by the vectors $u_k$ and $b_k$. It is convenient to introduce the hyperbolic rotation of the orthogonal pair $(u,b)$ to the orthogonal pair $(w,c)$:
\begin{align}
\begin{cases}
w_i = u_i \cosh\alpha  + b_i \sinh \alpha \\
c_i = u_i \sinh\alpha +b_i \cosh\alpha
\end{cases}
\alpha\neq 0 \label{rotalpha}
\end{align}
Then: $w^2=-1$, $c_kw^k=0$, $c^kc_k=1$, $u_iu_j-b_ib_j = w_iw_j-c_ic_j$. 
The choice that $w $ has a component parallel to $u$ is not a limitation: if $w $ exists, also $-w $ is time-like torse-forming with scalar field $-\lambda $. 

\begin{prop}
The only possible hyperbolic rotation is
\begin{align}
 \tanh\alpha = - \frac{2R_{ij}u^ib^j}{R_{ij}(u^iu^j+b^ib^j)} 
\label{angle}
\end{align}
\begin{proof}
Contraction of \eqref{ricci1} with $u^i$ and of \eqref{ricci2} with $w^i$ give:
$R_{ij}u^iu^j=R_{ij}w^iw^j=-\xi$. Then:
$0=R_{ij}(w^iw^j - u^iu^j)=\sinh\alpha [R_{ij} (u^iu^j+b^ib^j)\sinh\alpha
+2 (R_{ij} u^ib^j) \cosh\alpha]$. If $\alpha \neq 0$, the result is obtained.
\end{proof}
\end{prop}
\noindent

Let us write the condition \eqref{omega} in terms of the hyperbolic components:
\begin{align*}
& \nabla_i(u_j\cosh\alpha  +b_j \sinh\alpha ) \\
&=\lambda [ u_iu_j \cosh^2\alpha+  b_ib_j \sinh^2\alpha+
(u_ib_j + u_jb_i)\cosh\alpha\sinh\alpha  +  g_{ij} ]
\end{align*}
\begin{align*}
& (\nabla_i \alpha )(u_j \sinh\alpha +b_j \cosh\alpha )+\varphi \cosh\alpha (u_iu_j+g_{ij})+(\nabla_i b_j )\sinh\alpha  \\
&=\lambda [u_iu_j\cosh^2\alpha  +  b_ib_j \sinh^2\alpha +
 (u_ib_j + u_jb_i) \cosh\alpha\sinh\alpha+  g_{ij} ]
\end{align*}
Contraction with $u^j$, and the hypothesis $\sinh\alpha \neq 0$ give:
$$ \nabla_i\alpha = \lambda (u_i \sinh\alpha +b_i \cosh\alpha) -\varphi  b_i $$
Insertion in the previous equation gives, after simple algebra,
$$ \sinh\alpha (\nabla_i b_j)= (\lambda -\varphi \cosh\alpha) (u_i u_j + g_{ij}- b_ib_j) +\varphi b_i u_j \sinh\alpha $$
The trace of the equation gives the expression \eqref{lambda} for the parameter $\lambda $ and, if $\sinh\alpha \neq 0$,  the equation \eqref{condition}.


\begin{prop}
Condition \eqref{condition} is equivalent to the requirement that the space submanifold ($M^*,g^* $)
admits a unit vector $n^*_\mu (x)$ such that
\begin{align}
\nabla^*_\mu n^*_\nu = \frac{\nabla^*_\rho n^{*\rho}}{n-2} (g^*_{\mu\nu} - n^*_\mu n^*_\nu) + n^*_\mu  n^{*\prime}_\nu
 \label{nstar}
\end{align}
where $n^{*\prime}_\nu = n^{*\rho}\nabla^*_\rho n^*_\nu$.
\begin{proof}
In the comoving frame where $u^0=1$ (and $b^0=0$) the normalization $b^kb_k=1$ and the conditions \eqref{condition} become:
\begin{align*}
& a^{-2} g^{*\mu\nu} b_\mu b_\nu=1\\
& \partial_0 b_\mu - \Gamma_{0\mu}^\nu b_\nu =0\\
& \partial_\mu b_\nu - \Gamma_{\mu\nu}^\rho b_\rho = \tfrac{1}{n-2}(a^2 g^*_{\mu\nu} -b_\mu b_\nu) \frac{1}{a^2} g^{*\rho\sigma} (\partial_\rho b_\sigma -  \Gamma_{\rho\sigma}^\tau b_\tau )
\end{align*}
The second equation is $\partial_t (b_\mu/a)=0$. Then, the vector $n^*_\mu = b_\mu /a$ is normalized and independent
of time (it is a vector field of $M^*$). The last equation, with some algebra and use of the Christoffel symbols
in \cite{ManMol2017} becomes
\begin{align*}
\nabla^*_\mu n^*_\nu = \frac{\nabla^*_\rho n^{*\rho}}{n-2} (g^*_{\mu\nu} - n^*_\mu n^*_\nu) + n^*_\mu \frac{1}{a}(a_\nu -
n^*_\nu n^*_\rho a^\rho) 
\end{align*}
Contraction with $n^{*\mu}$ gives $ n^{*\prime}_\nu = \frac{1}{a} (a_\nu -
n^*_\nu n^*_\rho a^\rho)$, and \eqref{nstar} is obtained.
\end{proof}
\end{prop}

Eq.\eqref{nstar} coincides with eq.(7.9) by Coley and McManus \cite{ColeyMcM}, in $n=4$. 
The existence of the normalized vector $n^*_\mu $ with condition \eqref{nstar} (i.e. shear and vorticity free, but not geodesic) implies that ($M^*,g^* $) is a doubly twisted manifold, i.e. there are space coordinates and functions $f_1$, $f_2$  such that
$$g^*_{\mu\nu}(x) dx^\mu dx^\nu =   f_1(x)^2 (dx^1)^2 + f_2(x)^2 ds^2 $$
where $ds^2 $ only refers to coordinates $x^2, ...,x^{n-1}$ (see Table 1 in Borowiec and Wojnar \cite{Borowiec} and Corollary 1 in Ferrando et al. \cite{Ferrando}). 
In particular, the space manifold $(M^*,g^*) $ is twisted if and only if the vector fields $a_\mu $ and $n^*_\mu $ are also 
parallel.\\

In a twisted manifold, the general form of the Ricci tensor is \cite{ManMol2017}:
\begin{align}
R_{ij} = \frac{R-n\xi}{n-1}u_iu_j + \frac{R-\xi}{n-1}g_{ij}+(n-2)v (u_ib_j+u_jb_i) - (n-2) E_{ij} \label{RicciTW}
\end{align}
where $v=b^k\nabla_k \varphi $.  If another torse-forming vector $w_i$ exists, eq.\eqref{rotalpha}, 
the same Ricci tensor is: 
$$
R_{ij} = \frac{R-n\xi}{n-1}w_iw_j + \frac{R-\xi}{n-1}g_{ij}+(n-2)v' (w_ic_j+w_jc_i) - (n-2) E'_{ij}$$
where $E'_{ij}=w^r w^s C_{rijs}$ and $v' = c^k\nabla_k \lambda $. 

\begin{lem} $v'=-v$.
\begin{proof}
Contract of Eq.\eqref{13a} with $c^k$ and use $c^k w_k=0$:
$$c^k u_k [w^i\nabla_i\varphi +
w^i u_i (u^j \nabla_j\varphi) ] =[(u^iw_i)^2 -1] v'   $$
It is $c^ku_k =-\sinh\alpha \neq 0$, $w^ku_k=-\cosh\alpha $. Then:
$$(\cosh\alpha u^i + \sinh\alpha b^i) \nabla_i\varphi  -\cosh\alpha (u^j\nabla_j \varphi )
 =- \sinh\alpha v'   $$
 Simplify and use  $b^i\nabla_i\varphi =v$.
\end{proof}
\end{lem}

\begin{prop}
If $u_i$ and $w_i$ are non-collinear torse-forming vector fields, then the Weyl tensor $C_{jklm}$ has
the constraint
\begin{align}
(u^ru^s + b^rb^s) C_{rijs} =  (u_iu_j + b_ib_j) (b^kb^l E_{kl}) \label{Weyln}
\end{align}
where $E_{ij} =u^ru^s C_{rijs}$ is the electric component of the Weyl tensor,
with the properties $E_{ij}u^i=0$ and $E^i{}_i=0$.
\begin{proof}
Subtraction of the two expressions of the Ricci tensor and use of \eqref{rotalpha} with $\alpha \neq 0$ give:
\begin{align*} 
 \left[\frac{R-n\xi}{(n-1)(n-2)} \sinh\alpha-2v\cosh\alpha \right] [\sinh\alpha (u_iu_j+b_ib_j) +\cosh\alpha
(u_ib_j+u_jb_i)]\\
= \sinh\alpha [\sinh\alpha(u^ru^s+b^rb^s) +\cosh\alpha (u^rb^s+u^sb^r)]C_{rijs} 
\end{align*}
Contraction with $u^iu^j$ gives:
\begin{align*}
\frac{R-n\xi}{(n-1)(n-2)} \sinh\alpha-2v\cosh\alpha =  \sinh\alpha \, (b^rb^s E_{rs} )
\end{align*}
Then:
\begin{align}
& \sinh\alpha [(u_iu_j+b_ib_j) (b^rb^s E_{rs}) -(u^ru^s+b^rb^s) C_{rijs}] \label{eq24}\\
& =- \cosh\alpha [(u_ib_j+u_jb_i)(b^rb^s E_{rs}) - (u^rb^s+u^sb^r)C_{rijs} ] \nonumber
\end{align}
A torse-forming vector field has the property of being ``Weyl-compatible'' \cite{ManMolWeyl}: 
$$(u_iC_{jklm}+u_j C_{kilm}+ u_k C_{ijlm})u^m=0.$$
It implies $ C_{jklm}u^m = u_k E_{jl} - u_j E_{kl} $. Then
$$ (u^rb^s+u^sb^r)C_{rijs} = - b^s C_{jsir} u^r + b^r C_{rijs}u^s = (u_i E_{js}+u_j E_{is})b^s $$
Eq.\eqref{eq24} becomes:
\begin{align}
& \sinh\alpha [(u_iu_j+b_ib_j) (b^rb^s E_{rs}) -(u^ru^s+b^rb^s) C_{rijs}] \label{eq25}\\
& =- \cosh\alpha [(u_ib_j+u_jb_i)(b^rb^s E_{rs}) -   (u_i E_{js}+u_j E_{is})b^s] \nonumber
\end{align}
Contraction with $u^i$:
\begin{align}
\sinh\alpha [u_j (b^rb^s E_{rs}) +b^rb^s u^iC_{rijs}]  = - \cosh\alpha [b_j(b^rb^s E_{rs})  - E_{js}b^s ] \nonumber
\end{align}
The left-hand side of the equation is zero: $b^rb^s C_{rijs} u^i = b^rb^s C_{jsri} u^i = b^rb^s(u_sE_{jr}-u_j E_{sr}) =
-u_j (b^rb^s E_{rs})$. Then $b^j$ is eigenvector of $E_{js}$:
\begin{align}
E_{js} b^s = b_j (b^rb^s E_{rs}) \label{Eb}
\end{align}
and the right hand side of eq.\eqref{eq25} is zero.
\end{proof}
\end{prop}

\section{$n=4$, the two-fluid picture}
We show that in a space-time of dimension $n=4$ the presence of two torse-forming vectors specifies, via the Einstein equations, a stress-energy tensor that describes a mixture of two perfect fluids, with velocities $u_i$ and $u_i'$,
studied by Coley and McManus \cite{ColeyMcM}.

In $n=4$, as a consequence of Lovelock's identity \cite{LovRund}, the Weyl tensor is fully determined by its electric component $E_{ij}=u^ru^sC_{rijs}$:
$$ C_{ijkl} = 2(u_iu_l E_{jk} -u_i u_k E_{jl} +u_j u_k E_{il}-u_ju_l E_{ik})+
g_{il} E_{jk} -g_{ik} E_{jl} +g_{jk} E_{il}-g_{jl} E_{ik} $$
Contraction with $b^ib^l$ and use of \eqref{Eb} give:
$$ b^ib^l C_{ijkl} = (2u_j u_k -2b_j b_k +g_{jk})E_{rs} b^rb^s+E_{jk}  $$
Then \eqref{Weyln} gives $E_{ij}$ in terms of $b_i$, $h_{ij}=u_iu_j+g_{ij}$ and a scalar:
\begin{align}
2 E_{ij}= 3\left (b_ib_j - \tfrac{1}{3}h_{ij})(E_{rs} b^rb^s \right ). \label{Weyl4}
\end{align}
The Ricci tensor \eqref{RicciTW} becomes:
\begin{align}
R_{ij} = \tfrac{1}{3}(R-4\xi)u_iu_j + \tfrac{1}{3}(R-\xi ) g_{ij}+2 v (u_ib_j+u_jb_i)\\
 -  3\left (b_ib_j - \tfrac{1}{3}h_{ij} \right ) \nonumber
(E_{rs} b^rb^s )
\end{align}
Einstein's equations $R_{ij} -\frac{1}{2}R g_{ij}= T_{ij}$ give the corresponding energy-momentum tensor (in units that absorb the gravitational constant):
\begin{align*}
T_{ij} = \tfrac{1}{3}(R-4\xi )u_iu_j -\tfrac{1}{6}(R+2\xi) g_{ij}+2 v (u_ib_j+u_jb_i)\\
 -  3\left (b_ib_j - \tfrac{1}{3}h_{ij} \right ) 
(E_{rs} b^rb^s )
\end{align*}
The tensor, besides the perfect fluid-like term, contains a current term with vector $2vb_i$ orthogonal to the velocity,
and a peculiar stress tensor. 
This expression describes a mixture of two perfect fluids \cite{ColeyMcM}.\\
Consider two perfect fluids with velocities $u_i$ and $u'_i = u_i \cosh\theta + t_i \sinh\theta $, where the tilt angle
$\theta $ and the space-like unit vector $t_i$ are yet unspecified:
\begin{align*}
&T^{(2)}_{ij} =(p_1+\mu_1) u_i u_j +p_1 g_{ij} + (p_2+\mu_2)u'_i u'_j + p_2 g_{ij} \\
&= [(p_1+\mu_1) +(p_2+\mu_2)(1+\tfrac{4}{3}\sinh^2\theta )] u_i u_j + [p_1+p_2+ \tfrac{1}{3}(p_2+\mu_2)
\sinh^2\theta ]  g_{ij}\\
&+(p_2+\mu_2)\sinh\theta\cosh\theta (u_i t_j+u_j t_i) + (p_2+\mu_2) \sinh^2\theta ( t_i t_j -\tfrac{1}{3} h_{ij})
\end{align*}
If we equate $T_{ij}$ and $T_{ij}^{(2)}$, unicity of the 
decompositions with respect to the velocity field $u_i$, gives $t_i=b_i$ (up to a sign) and, with little algebra:
\begin{gather*}
 3(p_1+\mu_1)+3(p_2+\mu_2) = R-4\xi +12 E_{rs}b^rb^s, \quad 6(p_1+p_2)=-R-2\xi +6E_{rs}b^rb^s \\
(p_2+\mu_2)\sinh\theta\cosh\theta =2v, \quad   (p_2+\mu_2) \sinh^2\theta  = -3E_{rs}b^rb^s
\end{gather*}
The last two equations give the tilt angle between the fluid velocities $u$ and $u'$, while
the tilt angle bewteen the torse-forming vectors $u$ and $w$ is eq.\eqref{angle}: 
$$\tanh\theta =-\frac{3E_{rs}b^r b^s}{2v}, \quad \tanh \alpha = \frac{12 v}{R-4\xi -6 E_{rs}b^r b^s} $$
Thus $u'$ and $w$ are, in general, different time-like vectors. By expressing $b_i$ in terms of $u_i$ and $w_i$ we obtain:
$$ u'_i = u_i \frac{\sinh (\alpha -\theta )}{\sinh\alpha} - w_i \frac{\sinh\theta}{\sinh\alpha} $$
It turns out that $u'_i$ coincides with $w_i$ if $p_1+\mu_1=p_2+\mu_2$.

\section*{Appendix}
We report from \cite{ManMol2017} the Christoffel symbols, and the components of the Riemann and Ricci tensors 
for the metric \eqref{TW} of twisted space-times.\\ ($i,j,k,...=0,1,...,n-1; \, \mu,\nu,\rho,...=1,2, ... , n-1$).

{\quad}\\ \noindent
{\sf Christoffel symbols}: $\Gamma_{ij}^k = \Gamma_{ji}^k = \tfrac{1}{2}g^{km} (\partial_i g_{jm} +\partial_j g_{im} -\partial_m g_{ij})$. 
\begin{gather}
\Gamma_{i,0}^0=0,\quad  \Gamma_{0,0}^k=0, \quad \Gamma^\rho_{\mu,0} = (\dot a/a ) \delta^\rho_\mu, \quad 
\Gamma^0_{\mu,\nu} = a\dot a g^*_{\mu\nu}, \\
\Gamma^\rho_{\mu,\nu} = \Gamma^{*\rho}_{\mu,\nu} + (a_\nu/a)  \delta^\rho_\mu + (a_\mu/a)  \delta^\rho_\nu - (a^\rho/a)   g^*_{\mu\nu} 
\end{gather}
where $\dot a =\partial_t a$, $a_\mu = \partial_\mu a$ and $a^\mu = g^{*\mu\nu} a_\nu $.\\

\noindent
{\sf Riemann tensor}: $R_{jkl}{}^m = -\partial_j \Gamma^m_{k,l} + \partial_k \Gamma^m_{j,l} + \Gamma_{j,l}^p\Gamma^m_{kp} - \Gamma_{k,l}^p
\Gamma_{jp}^m $
\begin{align}
R_{\mu 0\rho}{}^0 =& (a\ddot a) g^*_{\mu\rho} \\
 R_{\mu\nu\rho}{}^0 =& g^*_{\mu\rho} (a \partial_\nu \dot a - \dot a a_\nu) - g^*_{\nu\rho} (a \partial_\mu \dot a - \dot a a_\mu ) \\
R_{\mu\nu\rho}{}^\sigma =& \, R^*_{\mu\nu\rho}{}^\sigma + ({\dot a}^2 - \frac{a^\lambda a_\lambda}{a^2} ) (g^*_{\mu\rho}\delta^\sigma_\nu - g^*_{\nu\rho}\delta^\sigma_\mu)\\
&+\frac{2}{a^2} ( a^\sigma a_\nu g^*_{\mu\rho} - a^\sigma a_\mu g^*_{\nu\rho} + a_\mu a_\rho \delta^\sigma_\nu - 
a_\nu a_\rho \delta^\sigma_\mu )  \nonumber \\
&+\frac{1}{a} \left[ \nabla^*_\mu (a^\sigma g^*_{\nu\rho} - a_\rho  \delta^\sigma_\nu)  - \nabla^*_\nu ( a^\sigma g^*_{\mu\rho} - a_\rho  \delta^\sigma_\mu)
\right ] \nonumber
\end{align}

\noindent
{\sf Ricci tensor}: $R_{jl} = R_{jkl}{}^k $
\begin{align}
R_{00}  =& -(n-1) (\ddot a / a) \\
R_{\mu 0} =& -(n-2)\partial_\mu (\dot a / a) \\
R_{\mu\nu} =& R^*_{\mu\nu} + g^*_{\mu\nu} [(n-2){\dot a}^2 + a\ddot a] +2(n-3)\frac{a_\mu a_\nu}{a^2} -(n-4) \frac{a^\sigma a_\sigma }{a^2} g^*_{\mu\nu}\\
&-(n-3) \frac{1}{a}\nabla^*_\mu a_\nu  - \frac{1}{a} g^*_{\mu\nu} \nabla^*_\sigma a^\sigma   \nonumber
\end{align} 

\noindent
{\sf Curvature scalar}: $R=R^k{}_k $
\begin{align}
R=&\frac{R^*}{a^2} +(n-1) \left [(n-2)\frac{{\dot a}^2}{a^2} + 2\frac{\ddot a}{a}\right ] \\
&- (n-2)\left[ (n-5)\frac{a^\sigma a_\sigma }{a^4} +2\frac{\nabla^*_\sigma a^\sigma  }{a^3} \right ]\nonumber
\end{align}

\vfill

\begin{thebibliography}{99}
%
%
\bibitem{Borowiec}
A.~Borowiec and A.~Wojnar, {\em Geometry of almost-product Lorentzian manifolds and relativistic observer},
arXiv:1302.1846 [gr-qc].
%
\bibitem{Chen1979}
B.-Y.~Chen, {\em Totally umbilical submanifolds}, Soochow J. Math. {\bf 5}  (1979), 9--37.
%
\bibitem{Chen2017}
B.-Y.~Chen, Differential geometry of warped product manifolds and submanifolds, World Scientific (2017).
%
\bibitem{Chen2014}
B-Y~Chen, {\em A simple characterization of
generalized Robertson-Walker space-times}, Gen. Relativ. Gravit. {\bf 46} (2014) 1833, 5 pp.
%
\bibitem{ColeyMcM}
A.~A.~Coley and D.~J.~McManus, {\em On space-times admitting shear-free, irrotational, geodesic time-like congruences},
Class. Quantum Grav. {\bf 11} (1994) 1261--1282.
%
\bibitem{ColeyMcM96}
A.~A.~Coley and D.~J.~McManus, {\em New slant on tilted cosmology}, Phys. Rev. D {\bf 54} (10) (1996) 6095--6100.
%
\bibitem{Ferrando}
J.~ J.~Ferrando, J.~A.~Morales and M.~Portilla, {\em Inhomogeneons space-times admitting isotropic radiation: vorticity-free case}, Phys. Rev. D {\bf 46} (2) (1992) 578--584.
%
\bibitem{LovRund}
D.~Lovelock and H.~Rund, Tensors, differential forms, and variational principles, Dover reprint (1989).
%
\bibitem{ManMolGRW}
C.~A.~Mantica and L.~G.~Molinari, {\em Generalized Robertson-Walker space-times, a survey}, 
Int. J. Geom. Meth. Mod. Phys. {\bf 14} n.3 (2017) 1730001, 27 pp.
%
\bibitem{ManMol2017}
C.~A.~Mantica and L.~G.~Molinari, {\em Twisted Lorentzian manifolds: a characterization 
with torse-forming time-like unit vectors}, Gen. Relativ. Gravit. {\bf 49} (2017) 51, 7 pp. 
%
\bibitem{ManMolWeyl}
C.~A.~Mantica and L.~G.~Molinari, {\em Weyl compatible tensors}, Int. J. Geom. Meth. Mod. Phys. {\bf 11} n.8 (2014) 1450070, 15 pp.
%
\bibitem{Ponge}
R.~Ponge and H.~Reckziegel, {\em Twisted products in pseudo-Riemannian geometry}, Geom. Dedicata {\bf 48} n.1
(1993) 15--25.
%
\end{thebibliography}
\end{document}